\documentclass[fleqn,twoside]{article}
\usepackage{espcrc2}

\usepackage{graphicx}
\usepackage[figuresright]{rotating}

\newcommand{\AmS}{{\protect\the\textfont2
  A\kern-.1667em\lower.5ex\hbox{M}\kern-.125emS}}

\title{A composite open resonator for compact X-ray source}

\author{E.G.Bessonov\address[MCSD]{P.N. Lebedev Physical Institute RAS,
117924, Leninsky prospect 53, Moscow, Russia}
R.M.Fechtchenko\addressmark[MCSD]}

\begin{document}

                       \begin {abstract}
The results of calculation of the finesse of a composite open resonator
for compact X-ray source are presented. The region of the resonator
parameters has been found where the finesse is changed unessentially.
\vspace{1pc}
\end{abstract}

\maketitle

               \section {Introduction}

Currently there has been considerable progress in the development of
super-reflection mirrors, high-finesse optical resonators
(super-resonators) and frequency stabilized high power cw and pulsed
mode-locked lasers. Optical resonators can be used as a photon storage
to accumulate a very high laser power. By this method, a finesse of the
optical resonators of $10 ^6$ can be achieved, which means the laser
power can be enhanced by this order \cite {chen}. The high reactive
power stored in the super-resonators can be used in gravitational
experiments in astrophysics, Laser-Electron Storage Rings \cite {chen}
- \cite {urakawa}, gravitational analogue of lasers (grasers) \cite
{bess2} or other applications.

The main reason for mirror degradation of high finesse dielectric open
resonators of free-electron lasers based on storage rings is the
deposition of chemical components on the mirror surface by X-ray
synchrotron radiation \cite {yasumoto}. The same degradation can occur
in Laser-Electron Storage Rings where the backward Compton scattering
of laser photons by electron beams is used for production of X-ray and
$\gamma$-ray radiation \cite {zhirong}, \cite {bess1}.  Possible
solution of the mirror degradation problem is using of a composite
resonator.  Below the results of calculation of the dependence of the
finesse of such a composite resonator on its parameters are presented
and analyzed.

          \section {Quality of the composite open resonator}

Let a composite open resonator consist of two mirrors (see Fig.1). The
reflectivity of the first mirror $M _1$ and the main part of the
second mirror $M _2$ are $R _1$. The second mirror has a circular
insertion at the axis of the resonator made from another material with
the reflectivity $R _2$. The radius of the insertion is "$a$", the rms
radius of the fundamental mode of the resonator at the surface of the
mirror is "$\sigma _{L m}$". Both mirrors can be dielectric multilayer
mirrors and the insertion part in one in two mirrors can be coated by
$Au$.

\vskip 45mm
\hskip 18mm
\begin{picture}(50,40)
\includegraphics[scale=0.53]{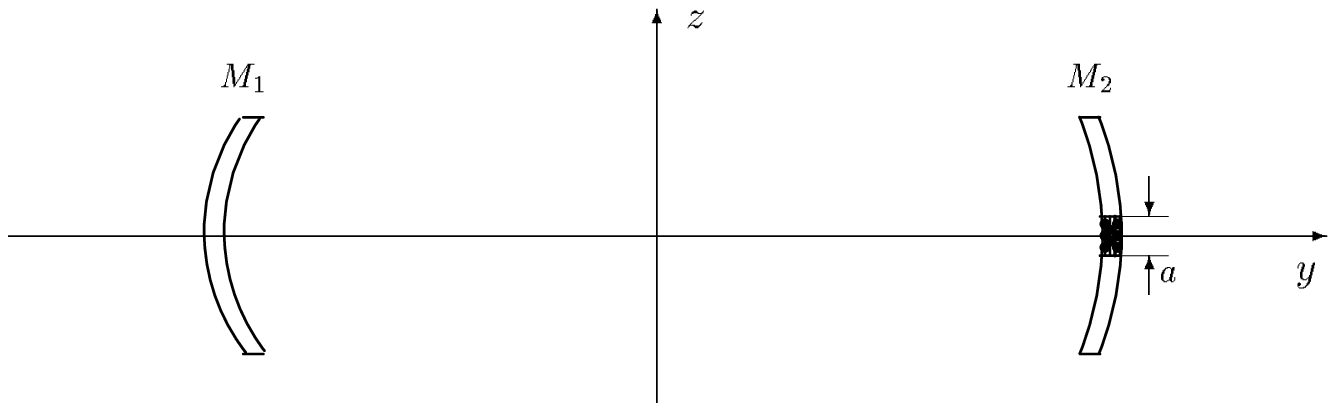}
\end{picture}
\vskip -25mm
{\small Fig 1: A composite open resonator.}

\vskip 10mm

The intensity of the Gaussian laser beam is distributed by the law

        \begin {equation} 
        I _L = {P _0 \over 2\pi \sigma _L ^2} e ^ {- {r ^2 \over 2
        \sigma _ {L} ^2}},
        \end {equation}
where $P _0 = \int I _LdS$ is the initial power of the laser beam
stored in the resonator; $dS$, the element of the area; $ \sigma _L =
\sigma _ {L \, 0} \sqrt {1 + s^2/ l _R^2} $, the dispersion of the
laser beam in a point $s$; $\sigma _ {L \, 0} = \sigma _L (s=0)$; the
point $s = 0 $ corresponds to the waist of the laser beam; $l _R =
4 \pi \sigma _ {L \,0} ^2 / \lambda _L$, the Rayleigh length; $\lambda
_L$, the laser wavelength.

In a composite resonator a part $P _1 = P _0 R _1$ of the laser power
is reflected by the first mirror. A part $(P _1 - \Delta P _1) R _1$ of
this power is reflected by external part of the second mirror and the
another part $\Delta P _1 R _2$ by the internal one, where $\Delta P _1
= P _1 [1 - \exp (-a ^2/2\sigma _{L m} ^2)]$ is the energy incident
upon the insertion part of the second mirror. The round trip reflected
power $P _2 = (P _1 - \Delta P _1)R _1 + \Delta P _1 R _2 $ can be
presented in the form

        \begin {equation} 
        P _2 = P _0 R _1 ^2 [1 - {R _1 - R _2 \over R _1} (1 - e ^{-{a
        ^2 / 2\sigma _{L m} ^2}})].
        \end {equation}

After $n$ round trips the laser light power can be presented in the
form

        $$P(t) \simeq P _0 ({P_2 \over P _0})^n = P _0 e ^{n \ln (P _2 /
        P _0)} = $$

        \begin {equation} 
        P _0 e ^{n \ln \{R _1 ^2 [1 - {R _1 - R _2 \over R _1}
        (1 - e ^{-{a ^2 / 2\sigma _{L m} ^2}})]\}},
       \end {equation}
where $n = ct/2L$; $L$, is the resonator length.

The finesse of the open resonator can be determined by the equation $F
= - (2\pi / T)[P /(\partial P/ \partial t)]$. In this case, according
to (3), the finesse of the composite resonator can be presented in the
form

        $$F = {- 2\pi \over \ln \{R _1 ^2 [1 - {R _1 - R _2 \over R _1}
        (1 - e ^{-{a ^2 / 2\sigma _{L m} ^2}})]\}} = $$

        \begin {equation} 
        F _0 {\ln R _1 ^2 \over \ln \{R _1 ^2 [1 - {R _1 - R _2 \over
        R _1} (1 - e ^{-{a ^2 / 2 \sigma _{L m} ^2}})]\}},
        \end {equation}
where $F _0 = F |_{R _1 = R _2} = - \pi / \ln R _1 \simeq \pi /(1 - R
_1)$.

According to (4), in the approximations $(1 - R _1)/(1 - R _2) \ll 1$,
$1 - R _1 \ll 1$, $1 - R _2 \ll 1$, $a/\sigma _{L m} \ll 1$, the
finesse of the resonator will be decreased $F _0/F$ times when the
radius

        \begin {equation} 
        a \simeq \sqrt {2} \sigma _{L m} \sqrt {{1 - R _1 \over 1 -
        R_2}({F _0 \over F} -1)}.  \end {equation}

{\it Example.} The relativistic factor of electrons $\gamma =10 ^3$,
$\lambda _L = 10$ $mkm$, $\sigma _{L \,0} = 50$ $mkm$, $L = 2m$, $F / F
_0 = 0.5$, $(1 - R _2) /(1 - R _1) = 10 ^2$.

In this case, according to (1), (5), the Rayleigh length $l _R = 3.14$
$mm$, the mode size $\sigma _{L m} = 15.9$ $mm$, the radius of the
X-ray beam at the mirrors $\sigma _{X-ray} \simeq L/2\gamma = 1$ $mm$,
the radius of the insertion $a = 0.1 \sqrt 2 \sigma _{L m} \simeq 2.25$
$mm$ $>\sigma _{X-ray}$. The magnification $M$, defined as the ratio of
the mode size on the mirrors to the mode size at the waist in the
resonator, must be chosen to be high ($\sim 10 ^2 \div 10 ^3$) as a
compromise between mirror degradation, resonator finesse and source
issues.

\section {Conclusion}

We have considered a composite open resonator. One of its mirrors has a
small circular insertion at the axis of the resonator made from another
much lower finesse material. It was shown that if the dimension of
the insertion is larger than the effective transverse dimension of the
X-ray beam and much less than the diameter of the fundamental mode
the mirror degradation time can be increased to a great extent and the
finesse of the resonator is decreased unessentially. Such conditions
can be fulfilled when the magnification $M \gg 1$.

This work was supported partly by the Russian Foundation for Basic
Research, Grant No 02-02-16209.

\end {document}